# "Programming" Quantum Hardware via Levenberg Marquardt Machine Learning

James E Steck
*Dept. of Aerospace Engineering*
Wichita State University
Wichita, KS USA

Nathan L. Thompson
*Dept. of Mathematics, Statistics and Physics*
Wichita State University
Wichita, KS USA

Elizabeth C Behrman
*Dept. of Mathematics, Statistics and Physics*
Wichita State University
Wichita, KS USA
ORCID. 0000-0002-6195-9122

*Abstract*—We present an improved method for quantum machine learning, using a modified Levenberg-Marquardt (LM) method. The LM method is a powerful hybrid reinforcement learning technique ideally suited to quantum machine learning as it only requires knowledge of the final measured output of the quantum computation, not intermediate quantum states which are generally not accessible. Thus, we are able with this method to achieve true online training of a quantum system to do a quantum calculation. We demonstrate this using a fundamentally non-classical calculation: estimating the entanglement of an unknown quantum state. Machine learning is demonstrated with results from simulations and runs on the IBM Qiskit hardware interface.

## I. Introduction

The term "quantum machine learning" is usually used to mean the use of classical machine learning algorithms for analysis of non-quantum data on a quantum computer. This is more properly called "quantum assisted machine learning." In contrast, we have developed *true* quantum machine learning: use of a quantum system as a quantum computer that learns a *quantum* task. The benefits of the machine learning approach are manifold: (1) it bypasses the algorithm construction problem, since the system itself learns to find the desired procedures; (2) it enables computations without breaking down the procedure into a sequence of steps (sequence of CNOT, Hadamard, rotations, etc.: its "building blocks", potentially increasing efficiency; (3) scaleup is easy, almost immediate; and (4) the multiple interconnectivity of the architecture means that the computations are robust both to noise and to decoherence.

We contrast our method with the "building block" strategy, which is the usual algorithmic approach, in which the procedure is formulated as a sequence of steps (quantum gates) from a universal set, e.g., a sequence of CNOT, Hadamard, and phase shift gates. This is of course exactly analogous to the way in which classical computing is usually done, as a series of logical gates operating on bits. But our method follows a different computing paradigm: that of distributed computing, which is the approach of biological and of artificial neural networks. Since the 1990s our research group has been investigating this different approach, a combination of quantum computing and artificial neural networks, as an alternative to the building block paradigm. With this approach, the quantum systems itself learns how to solve the problem, designing its own algorithm in a sense. Moreover, we (and others) have shown [1] [2] that not only does this obviate of the program design obstacle, but also gives us near-automatic scaling [3], robustness to noise and to decoherence [4], and speedup over classical learning [5] [6].

The basic idea is that a quantum system can itself act as a neural network: The state of the system at the initial time is the "input"; a measurement (observable) on the system at the final time is the "output"; the states of the system at intermediate times are the hidden layers of the network. If we know enough about the computation desired to be able to construct a comprehensive set of input-output pairs from which the net can generalize, then, we can use techniques of machine learning to bypass the algorithm-construction problem.

Entanglement is an intrinsically quantum mechanical property, essential for most quantum speedups. It can be calculated, given the state of the system (wavefunction or density matrix) for a two-qubit system, but if the state of the system is unknown, determination of the entanglement requires quantum tomography. And for systems larger than two qubits, determination of the entanglement is an NP-hard problem [7] Thus, entanglement estimation is a good example of a nontrivial, intrinsically *quantum mechanical, calculation* for which we have no general algorithm [8].

In previous work we succeeded in finding a time dependent Hamiltonian for a multiqubit system such that a chosen measurement at the final time gives a witness of the entanglement of the initial state of the quantum system [2] [3]. The "output" (result of the measurement of the witness at the final time) will change depending on the time evolution of the system, which is of course controlled by the Hamiltonian: by the tunneling amplitudes, the qubit biases, and the qubit-qubit coupling. Thus, we can consider these parameters to be the "weights" to be trained. We then use a quantum version [2] of backpropagation [9] to find optimal quantum parameters such that the desired mapping is achieved. (It should be noted that the method of quantum backprop has recently [10] [11] been rediscovered by several groups.) Full details are provided in [2]. From a training set of only four pure states, our quantum neural network successfully generalized the witness to large classes of states, mixed as well as pure [3]. Qualitatively, what we are doing is using machine learning techniques to find a "best" hyperplane to divide separable states from entangled ones, in the Hilbert space.

Of course, this method is necessarily "offline" training, since it is not possible to do quantum backpropagation without knowing the state of the system at intermediate times (in the hidden layers); quantum mechanically, this is impossible without collapsing the wavefunction and thereby destroying the superposition, which rather obviates the whole purpose of doing quantum computation. That is, quantum backpropagation can only be done on an (auxiliary) classical computer, simulating the quantum computer. This simulation will necessarily contain uncertainties and errors in modeling the behavior of the actual quantum computer. The results from offline quantum backpropagation, can, of course, be used as a good starting point for true online quantum learning, where this online learning is used to correct for uncertainty, noise, and decoherence in the actual hardware of the quantum computer. Here, we present such a method, port it to the IBM Qiskit system [12], and demonstrate its effectiveness. The next section introduces our machine learning method for *deep time quantum networks*, including our original quantum backprop method and our equivalent hardware compatible implementation of the results. In Section III we describe a learning model implementable on hardware, a parameter variation finite difference method, and show results on both Matlab simulation and Qiskit. In Section IV we develop our new model using Levenberg Marquardt, and again show and compare results using both Matlab and Qiskit. We conclude in Section V. This is an extension of work previously presented here [13].

## II. Machine Learning for Deep Time Quantum Networks

### A. Machine Learning in Simulation

A general quantum state is mathematically represented by its density matrix, $\rho$, whose time evolution obeys the Schrödinger equation

$$\frac{d\rho}{dt} = \frac{1}{i\hbar}[H, \rho] \qquad (1)$$

where H is the Hamiltonian and $\hbar$ is Planck's constant divided by $2\pi$. The formal solution of the equation is

$$\rho(t) = e^{i\mathcal{L}t}\rho(t_0) \qquad (2)$$

where $\mathcal{L}$ is the Liouville operator, defined as the commutator with the Hamiltonian in units of $\hbar$, $\mathcal{L} = \frac{1}{\hbar}[H, \dots]$. We can think of (1) as analogous to the equation for information propagation in a neural network, Output = FW*Input, where FW represents the action of the network as it acts on the input vector Input. The time evolution of the quantum system, given in (2), maps the initial state (input) to the final state (output) in much the same way. The mapping is accomplished by the exponential of the Liouville operator, $e^{i\mathcal{L}t}$. The parameters playing the role of the adjustable weights in the neural network are the parameters in the Hamiltonian that control the time evolution of the system: the physical interactions and fields in the quantum hardware, which can be specified as functions of time, just as, in the gate model, different gates are implemented in a given sequence. Because we want to be able physically to implement our method, we use not the final state of the system, $\rho(t)$, as our output, but instead a measure, M, applied to the quantum system at that final time, producing the output $O(t_f) = M(\rho(t_f))$. "Programming" this quantum computer, the act of finding the "program steps or algorithm", involves finding the parameter functions that yield the desired computation. We use machine learning to find the needed quantum algorithm. This means we learn the system parameters inside H to evolve in time initial (input) to target final (output) states; yielding a quantum system that accurately approximates a chosen function, such as logic gates, benchmark classification problems, or, since the time evolution is quantum mechanical, a quantum function like entanglement. If we think of the time evolution operator in terms of the Feynman path integral picture [14], the system can be seen as analogous to a deep neural network, yet quantum mechanical. That is, instantaneous values taken by the quantum system at intermediate times, which are integrated over, play the role of "virtual neurons" [2]. In fact, this system is a deep learning system, as the time dimension controls the propagation of information from the input to the output of the quantum system, and the depth is controlled by how finely the parameters are allowed to vary with time. We use the term "dynamic learning" to describe the process of adjusting the parameters in this differential equation describing the quantum dynamics of the quantum computer hardware. The real time evolution of a multi-qubit system can be treated as a neural network, because its evolution is a nonlinear function of the various adjustable parameters (weights) of the Hamiltonian.

We define a cost function, the Lagrangian L, to be minimized, as

$$L = \tfrac{1}{2}[d - O(t_f)]^2 + \int_{t_0}^{t_f} \lambda^\dagger(t)\left(\tfrac{\partial \rho}{\partial t} + i\hbar[H,\rho]\right)\gamma(t)dt \quad (3)$$

where the Lagrange multiplier vectors are $\lambda^\dagger$ and $\gamma$ (row and column, respectively), where d is the desired value, and where $O(t_f)$ is the output measure at the final time. Note that this will constrain the density matrix to satisfy the Schrodinger equation during the time interval. In the example application presented later in this paper, we define the output measure for our quantum system to be trained as a pairwise entanglement witness for qubits α and β [2] as

$$\langle O(t_f)\rangle = tr[\rho(t_f)\sigma_{z\alpha}\sigma_{z\beta}] \quad (4)$$

where tr stands for the trace of the matrix, the pointed brackets indicate the average or expectation value, and $\sigma_z$ is the usual Pauli matrix. That is, the output is the qubit-qubit correlation function at the final time. This measure is chosen for this calculation because entanglement is a kind of quantum correlation, so it makes sense to choose to map the entanglement of the initial state of the system to an experimental measure of correlation. To implement quantum backprop we take the first variation of L with respect to ρ, set it equal to zero, then integrate by parts to give the following equation which can be used to calculate the vector elements of the Lagrange multipliers ("error deltas" in neural network terminology) that will be used in the learning rule:

$$\lambda_i \tfrac{\partial \gamma_j}{\partial t} + \tfrac{\partial \lambda_j}{\partial t}\gamma_j - \tfrac{i}{\hbar}\sum_k \lambda_k H_{ki}\gamma_j + \tfrac{i}{\hbar}\sum_k \lambda_i H_{jk}\gamma_k = 0 \quad (5)$$

which is solved backward in time under the boundary conditions at final time $t_f$ given by $-[d - \langle O(t_f)\rangle]O_{ji} + \lambda_i(t_f)\gamma_j(t_f) = 0$. In optimization and optimal control this is called the co-state equation. The gradient descent rule to minimize L with respect to each network weight w (where w is one of the quantum parameters) is $w_{new} = w_{old} - \eta\tfrac{\partial L}{\partial w}$, where

$$\frac{\partial L}{\partial w} = \lambda^\dagger(t)\left(i\hbar\left[\frac{\partial H}{\partial w},\rho\right]\right)\gamma(t)$$

Because this technique uses the density matrix, it is applicable to any general state of the quantum system, pure or mixed. While this method works extremely well to train quantum systems in simulation, the gradient $\frac{\partial L}{\partial w}$ requires knowledge of the quantum state ρ, the density matrix, at each time t from $t_0$ to $t_f$. This makes this method not amenable to real time quantum hardware training, since measuring the quantum state at intermediate times collapses the quantum state and destroys the quantum mechanical computation [15]. In other words, quantum backprop can be run in simulation and the resulting approximate parameters executed on quantum hardware as we have done in [16], but training on the hardware itself cannot be accomplished. Also, because the H used in the above "off-line" machine learning does not exactly match the quantum hardware due to unknown or unmodeled dynamics and uncertainties in the physical system, the resulting calculations on the hardware will have some error. The "off-line" parameters can, however, be a starting point for further machine learning refinement on the hardware using the techniques described in the next sections.

*B. A Hardware Compatible Model for IBM Qiskit*

Implementing the pairwise entanglement witness in a hardware compatible model, such as IBM's Qiskit [12], requires some modifications. The Qiskit library utilizes a quantum gate model, so we must convert and restrict our more general Hamiltonian to a gate representation of that operator. The witness is constructed by first approximating the values of the tunneling, bias, and coupling parameters as piecewise constant, where the total evolution time is divided into 4 segments. These piecewise constant parameters are used to form the Hamiltonian for the time evolution operator, which is converted into a sequence of gates, a quantum circuit for the IBM Qiskit implementation. This circuit representation results in 20 independent weights $w_j$ for the entanglement witness. One of the time segments of the circuit is pictured in Fig. 1, where $R_y$ and $R_z$ are the rotations around the y and z axes, respectively, and the other circuit symbol represents the CNOT gate.

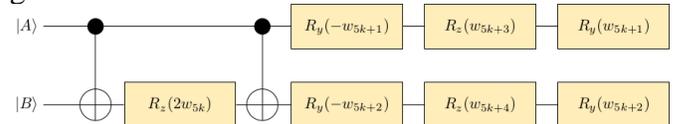

Fig. 1. The entanglement witness quantum circuit structure for a single time segment.

The structure from Fig. 1 is repeated for each time segment, with weights on each segment determined by the piecewise constant approximations to the continuous parameter functions. Full details of the gate representation and a comparison with the behavior of the continuous parameters can be found in [5].

### III. Finite Difference Gradient Descent Learning on Quantum Hardware

*A. Fourier Quantum Parameters for Simulations*

As described above, each of the quantum system parameters that serve as the quantum computer algorithm can vary with time. For learning in the hardware environment, each quantum parameter/weight w(t) is represented as a Fourier series expansion in time:

$$w(t) = w_0 + \sum_{j=1}^{n} S_j \sin\left(\frac{j\pi}{t_f}t\right) + C_j \cos\left(\frac{j\pi}{t_f}t\right) \quad (6)$$

where $t_f$ is the final time when the quantum system output measures are taken. This gives a limited population of Fourier coefficients for which to calculate the gradients needed for the LM learning algorithm described below. This is motivated by our results shown in [17] where offline quantum backpropagation is used to train general time varying quantum parameters. The functions *w(t)* were allowed to be any continuous function of time, but it was discovered that the resulting parameters have obvious simple frequency content. In that paper we showed that fitting the parameters with Fourier series for sine and cosine gave equivalent computing results.

Learning of each of the parameters is done via a parameter variation hybrid method which uses small variations of the Fourier coefficients to calculate the gradient of the output error which is then used in a straightforward gradient descent learning rule. For a given training pair in the training set, the quantum system is presented with the input, the system runs (with the current parameters calculated from the current Fourier coefficients) until the final time $t_f$ where the output is calculated via the output measures on the final state. The output is compared to the target value and an output error is calculated $E_{old}$. In the backpropagation method, this output error is then backpropagated via quantum backprop to calculate gradients at each time step. In the hybrid reinforcement learning method, the following happens.

Choosing a single parameter and Fourier coefficient in (6), this coefficient is varied by a small amount. For example, the new value parameter $w_0$ would be given by

$$w_{0,new} = w_0 + \Delta w_0. \tag{7}$$

The quantum system is again presented with the input; the system then runs with the parameters calculated using the modified Fourier coefficients; the output is calculated; an output error $E_{new}$ is calculated and compared to the error $E_{old}$; a gradient is calculated

$$\nabla E = \frac{E_{new} - E_{old}}{\Delta w_0} \tag{8}$$

and, finally, this gradient is used to update the parameter using a specified learning rate $\eta w_0$ via

$$w_0 = w_0 + \eta w_0 \nabla E. \tag{9}$$

This is repeated for all Fourier coefficients representing each quantum parameter, using the same input and target output. Each successive training pair is then processed in the same way until the entire list of training pairs is exhausted, constituting one epoch of training.

## B. Parameter Variation Finite Difference Gradients Learning Results

MATLAB® code implements the learning algorithm above and calls a MATLAB® simulation of the quantum system. Compared to the quantum backprop method, finite difference gradients, in simulation, takes about 25 times more computation time. The tunneling frequency is initialized to $2.5 \times 10^{-3}$ GHz, is varied by 0.02% to calculate the gradient and a learning rate of $2 \times 10^{-8}$ is used. The bias is initialized to $10^{-4}$ GHz, is varied by 0.02% to calculate the gradient and a learning rate of zero (not trained) is used. The qubit coupling matrix off-diagonal elements representing qubit-to-qubit coupling are initialized to $10^{-4}$ GHz, is varied by 0.02% to calculate the gradient and a learning rate of $4 \times 10^{-7}$ is used. The on-diagonal coupling of a qubit to itself is, of course, zero. The entanglement witness calculation described above is the quantum "program" to be learned. Three Fourier parameters in (6) are used, that is n = 3. Systems with 2, 3, 4 and 5 qubits are trained, using a method we call iterative staging or transfer learning [3], whereby knowledge about the smaller system is used to initialize training for the larger system A plot of the root-mean-square (RMS) error vs training epochs as well as plots of how each quantum parameter varies with time after training is complete have been presented previously in [13]. In this previous paper, the quantum backprop method is compared to the finite difference gradient methods. This current paper is focused on using a much faster and more robust Levenberg Marquardt method to learn the same entanglement witness algorithm.

## C. Finite Difference Gradient Descent Learning on IBM Qiskit

For finite difference gradients for Qiskit, the training process is very similar to the MATLAB® implementation, with necessary changes for the Qiskit system. First, one of the training states is evolved on the quantum system, then the measure chosen for the entanglement witness is applied to it giving an expectation value for the witness. Using the current weights, expectation values for each state in the training set are computed and subtracted from the target values to generate an RMS difference output error $E_{old}$. A single weight $w_j$ is adjusted by a small amount as in (7), and the output error is then computed with the modified $w_{j,new}$, yielding $E_{new}$. Equations (8) and (4) are used to update $w_j$ according to the specified learning rate η, and the process is repeated for each of the 20 weights and all 4 training pairs, constituting one epoch of training. Qiskit system initialization and training parameters are given in Table 1. Experimentation revealed that the system was most sensitive to changes in the tunneling, which is why it has a higher learning rate. Training was successful, but improvement stopped after approximately 2500 epochs where the RMS error oscillated near 0.02. Again, full results for finite difference training on IBM Qiskit have been presented in our previous paper [13].

Table 1. Qiskit Reinforcement Learning Initial Values

| Quantum Parameter | Initial Value | Perturbation | Learning Rate |
|---|---|---|---|
| Tunneling K | $2.0 \times 10^{-3}$ GHz | 0.02% | $10^{-2}$ |
| Bias ε | $1.0 \times 10^{-4}$ GHz | 0.02% | $10^{-3}$ |
| Coupling ζ | $1.0 \times 10^{-4}$ GHz | 0.02% | $10^{-3}$ |

IV. Levenberg Marquardt Learning for Quantum Hardware

A. Levenberg Marquardt Algorithm applied to quantum computing

Straightforward parameter variation finite difference gradients learning works and training converges, but the requirement of 2500 training epochs is untenable on near-term hardware. As such, we seek a more efficient learning scheme. A candidate is the Levenberg Marquardt (LM) algorithm [18] [19] for solving non-linear least-square problems. Strictly speaking, training the entanglement witness is not a least-square problem (our training set has only 4 elements in the 2-qubit case), but we will show that a LM-inspired weight update rule is nonetheless very effective and efficient.

The LM algorithm uses the learning rule

$$\delta w = - \left( J^T J + \lambda D^T D \right)^{-1} \nabla E \qquad (12)$$

to update each weight vector w, where λ is the damping factor, $D^T D$ is the scaling matrix, and $\nabla E$ is the error gradient. (The specifics of selecting the damping factor and scaling matrix are presented later in this section.) The Jacobian matrix J, comprised of gradients w.r.t. the quantum parameters, is given by

$$J = \begin{bmatrix} \frac{\partial O(x_1, w)}{\partial w_1} & \cdots & \frac{\partial O(x_1, w)}{\partial w_W} \\ \vdots & \ddots & \vdots \\ \frac{\partial O(x_N, w)}{\partial w_1} & \cdots & \frac{\partial O(x_N, w)}{\partial w_W} \end{bmatrix}. \qquad (13)$$

where $O(x_i, w)$ is the network output function evaluated at the $i^{th}$ input vector $x_i$ using the weights w with N and W being the total number of training inputs and weights, respectively.

For small λ, the update rule is similar to the Gauss-Newton algorithm, allowing larger steps when the error is decreasing rapidly. For larger λ, the algorithm pivots to be closer to gradient descent and makes smaller updates to the weights. This flexibility is the key to LM's efficacy, changing λ to adapt the step size and update method to respond to the needs of convergence: moving quickly through the parameter space where the error function is steep and slowly when near an error plateau and thereby finding small improvements. Our implementation is a modified LM algorithm following several suggestions in [20]. One epoch of training consists of the following:

1) Compute the Jacobian (13) with current weights w
2) Update the scaling matrix DTD and damping parameter λ
3) Calculate a potential update δw using (12), setting $w_{new}$ = w + δw
4) Find if RMS error has decreased with new weights, or if an acceptable uphill step is found
5) If neither condition in step 3 is satisfied, reject the update, increase λ, and return to step 2

6) For an accepted downhill or uphill step, set w = $w_{new}$ and decrease λ, ending the epoch

Partial derivatives in the Jacobian are computed using the parameter-shift rule [21]. The scaling matrix $D^TD$ serves the primary purpose of combating parameter evaporation [22], which is the tendency of the algorithm to push values to infinity when somewhat lost in the parameter space. Following [20], we choose $D^TD$ to be a diagonal matrix with entries equal to the largest diagonal entries of $J^TJ$ yet encountered in the algorithm, with a minimum value of $10^{-6}$. Updates to the damping factor may be done directly or indirectly; our results here use a direct method. Analyzing the eigenvalues of the approximate Hessian $J^TJ$, we note that there is a cluster on the order of $10^{-4}$ and the rest nearly vanish. Testing showed that direct adjustments to the damping factor within a couple of orders of magnitude of these values resulted in more consistent training. For the minimum and maximum Hessian eigenvalues $l_{min}$ and $l_{max}$ in this cluster, we establish a logarithmic scale that ranges $[l_{min}/10, 1/l_{max}]$ with 100 elements. Following the principle of "delayed gratification" [22], we move 10 steps down the scale when an update is accepted and move 1 step up after rejecting an update. The log scale is desirable, because it allows the damping factor to change more slowly when close to the top end of the range. Classically, λ is modified by a multiplicative factor [19], but this causes the damping factor to change too rapidly for our problem once it becomes large.

Occasionally, the algorithm will fail to find a suitable update prior to reaching the top of the damping factor range. This could be due to a plateau [23] in the cost (RMS error) function or due to noise in the measurements causing the algorithm to miss a step it could have taken at a particular λ. When this occurs, we recompute the range for λ using the current Hessian and set the damping factor to be equal to the minimum value. Doing this provides the LM-algorithm the opportunity to randomly search the parameter space due to the stochastic nature of the quantum hardware measurements used to compute the Jacobian. To allow further exploration of the parameter space, we allow for uphill steps following the criterion suggested in [20]. This approach will accept an uphill step of the form

$$(1 - \beta)E_{i+1} \leq \min(E_1, E_2, ..., E_i) \quad (14)$$

where $E_i$ is the error in the $i^{th}$ iteration and

$$\beta = \cos(\delta w_{new}, \delta w_{old}) \quad (15)$$

is the cosine of the angle between the vectors formed by the proposed and last accepted weight updates, respectively. Criterion (14) checks if the angle between those update vectors is acute and will accept an uphill step within a tolerance. The longer training goes, the smaller an uphill step must be to be accepted. This feature allows the training to more easily move out of shallow local minima in the cost function.

B. Levenberg Marquardt training: MATLAB simulation results

The LM algorithm was added as a $3^{rd}$ training option in the backprop and finite difference MATLAB code. For any method, training in simulation for more than 5 qubits resulted in code runtimes longer than several days on a Windows PC. With the efficiency of the LM algorithm, 6 qubits could be accomplished on a PC. Beyond 6 qubits, the code execution was moved to a XSEDE cluster computer [24]. The training pair loop was distributed to a pool of 36 cores running on a node with a GPU via a MATLAB "parfor" statement replacing the for loop. After the Jacobian was completed for all training pairs, the LM matrix operations were done entirely on the GPU via GPU array functions and then gathered back from the GPU for the LM parameter (weight) updates. Training for the 2-qubit case was completed first. The RMS-vs-epoch, LM Lambda Parameter -vs- epoch are shown in Figs. 2-7. RNP is a noise parameter that is currently set to 0. To train the larger systems, we applied the trained 2-qubit parameters to initialize training for the 3-qubit system, then the trained 3-qubit parameters to initialize the 4-qubit system, and so on. We used this transfer learning successively to train all the way up to 8 qubits. Table 2 contains information for each qubit case. As the number of qubits increases, the transfer learning is more effective, and a significantly fewer number of epochs are needed as the parameters needed for the entanglement witness converge to constant values as shown in Figs. 8-9.

Table 2 Transfer learning for increasing numbers of qubits

| #qubits | #Training Pairs | #epochs | Start RMS | Finish RMS |
|---|---|---|---|---|
| 2 | 4 | 20 | .5439 | .0024 |
| 3 | 12 | 20 | .3108 | .0066 |
| 4 | 24 | 20 | .0569 | .0045 |
| 5 | 40 | 10 | .0263 | .0058 |
| 6 | 60 | 10 | .0204 | .0028 |
| 7 | 84 | 10 | .0166 | .0087 |
| 8 | 112 | 10 | .0160 | .0110 |

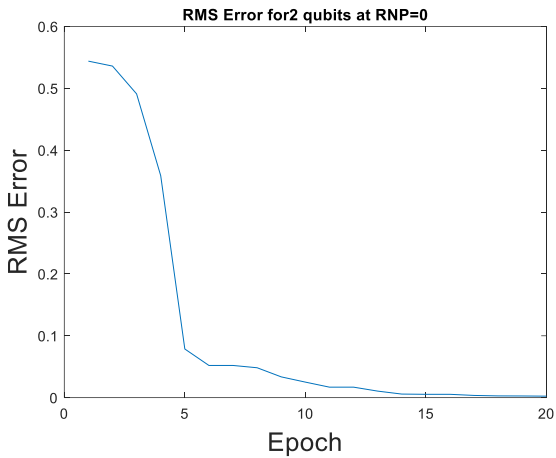

Fig. 2. RMS error vs epoch for 2-qubit entanglement witness training using the Levenberg-Marquardt method in a MATLAB Simulation.

*C. Levenberg Marquardt Qiskit training results*

Training with LM in Qiskit is markedly faster than gradient descent learning for the 2-qubit case. Fig. 10 shows training converged in approximately 30 epochs, nearly 100 times faster than the previous method shown in [17]. (Flat sections in the figure represent epoch(s) where the λ parameter reached the maximum of its range and triggered a recalculation of the quantum computations.) In addition, the RMS error reduced to a lower value, on the order of $10^{-4}$, in the Qiskit simulator; that is, the training with LM is both faster and better. This improvement is likely linked to the way LM deals with the ubiquitous problem [23] of barren plateaus and makes the LM method a potential major upgrade for online training on quantum hardware.

Moving to higher qubit values has been more of a challenge for the LM method. Training times for the 3-qubit case were longer, even after beginning with the trained 2-qubit values. That the 3-qubit case required extended training time is not unexpected, since the entanglement witness must learn to account for symmetries not present in the 2-qubit case [3]. Even so, Fig. 11 shows the 3-qubit cases train in less time than the method used in [13], which required 2500 epochs to the 300 needed to achieve the same 0.02 RMS error level. This result is encouraging as it is both a larger system and trained in nearly an order of magnitude fewer epochs. Scaling up to the 4-qubit case presents more difficulties, shown in Fig. 12. Training requires 600 epochs and improvements to the RMS error stalls out at approximately 0.05. We are examining our methodology and to find ways to improve both the 3-qubit and 4-qubit cases and then pushing to even higher qubit systems.

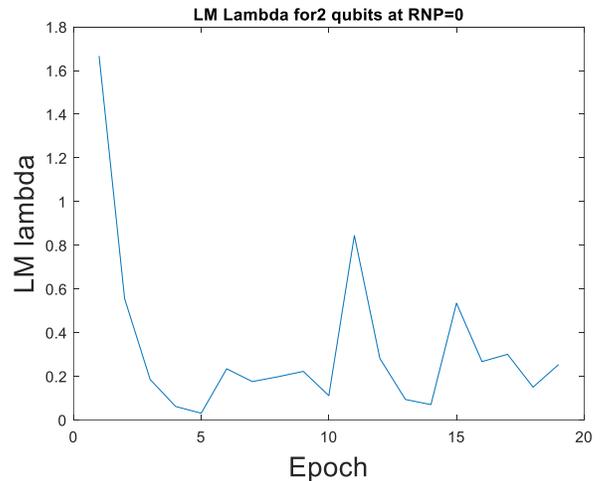

Fig. 3. LM Lambda vs epoch for 2-qubit entanglement witness training using the Levenberg-Marquardt method in a MATLAB Simulation.

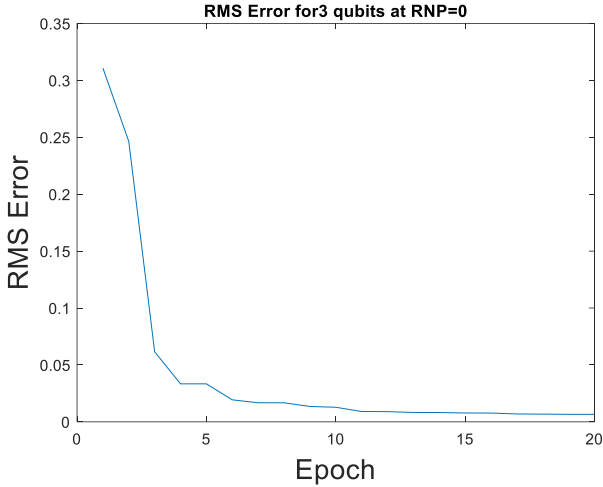

Fig. 4. RMS error vs epoch for 3-qubit entanglement witness training using the Levenberg-Marquardt method in a MATLAB Simulation.

## V. Discussions and conclusions

The major contribution of this paper is the demonstration of the feasibility of true online training of a quantum system to do a quantum calculation. It is a well-known theorem that a very small set of gates (e.g., the set {H, T, S, CNOT}) is universal. This means that any N-qubit unitary operation can be approximated to an arbitrary precision by a sequence of gates from that set. But there are many calculations we might like to do, for which we do not know an optimal sequence to use, or even, perhaps, any sequence to use. And there are many questions we might want to answer for which we do not even have a unitary, that is, an algorithm. Calculation of entanglement of an N-qubit system is a good example of such a question: we do not have a general closed form solution, much less know an optimal set of measurements to make on a system whose density matrix is unknown, to determine its entanglement.

Quantum machine learning methods like the ones used here are systematic methods for dealing with these problems. Here we show that they are in fact directly implementable on existing hardware. Our iterative staging technique makes scale-up relatively easy, as most of the training for a system of N qubits has already been accomplished in the system for (N-1) qubits. Additionally, this does not bias the results, as this entanglement witness performs very well when tested on larger systems, even in the face of noise [17] [4]. And while training on actual quantum hardware does prove somewhat more challenging, that is all the more reason for a machine learning approach. Any physical implementation features sources of error that in general are unknown (interactions, flaws, incomplete and damaged data). With machine learning we can deal with all these problems automatically.

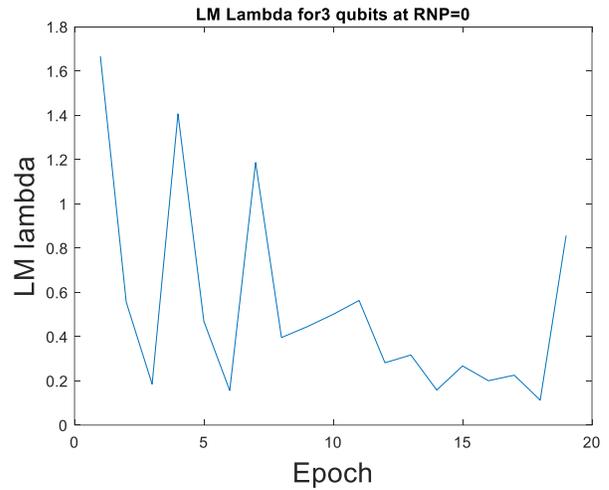

Fig. 5. LM Lambda vs epoch for 3-qubit entanglement witness training using the Levenberg-Marquardt method in a MATLAB Simulation.

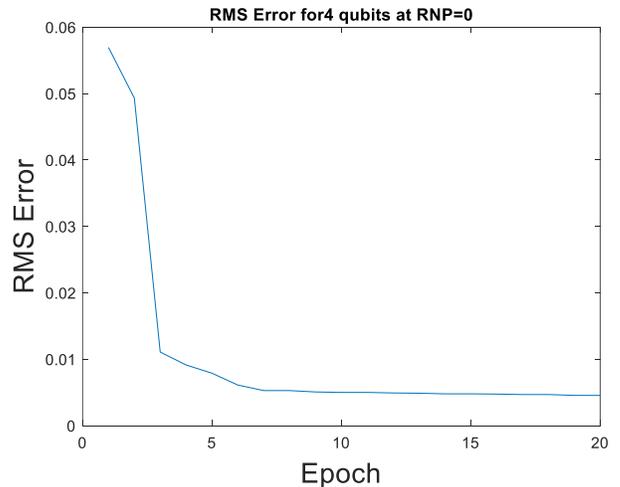

Fig. 6. RMS error vs epoch for 4-qubit entanglement witness training using the Levenberg-Marquardt method in a MATLAB Simulation.

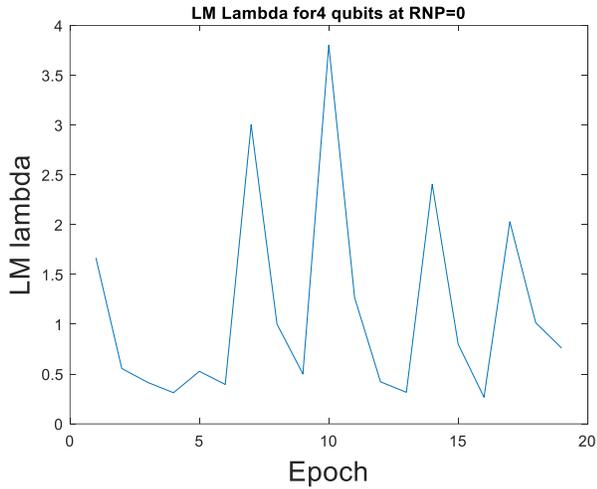

Fig. 7. LM Lambda vs epoch for 4-qubit entanglement witness training using the Levenberg-Marquardt method in a MATLAB Simulation.

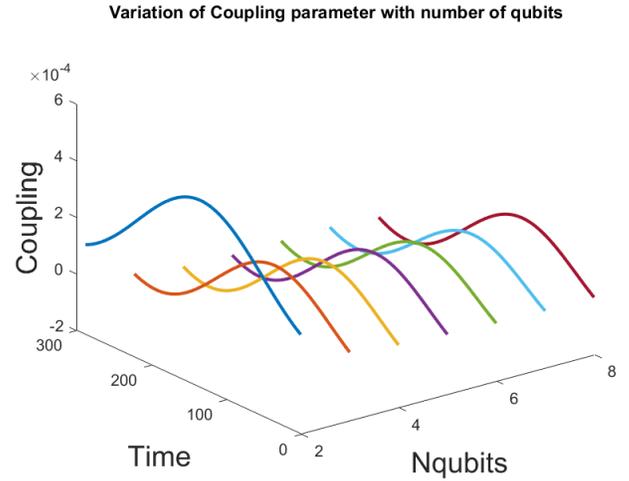

Fig. 9. Convergence of Coupling Parameter vs time as the number of qubits increases for the entanglement witness LM training in MATLAB Simulation

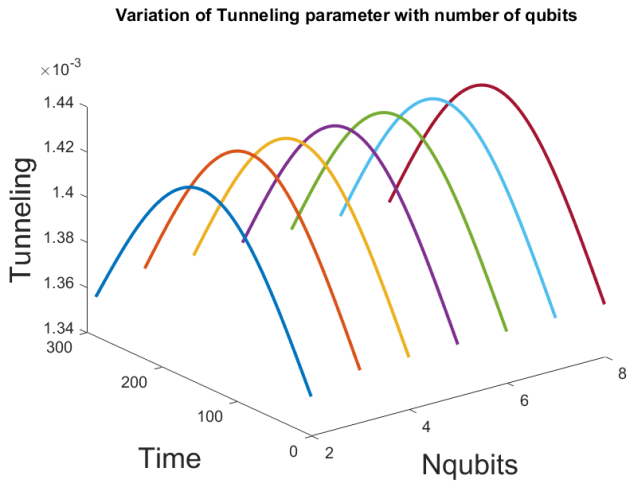

Fig. 8. Convergence of Tunneling Parameter vs time as the number of qubits increases for the entanglement witness LM training in MATLAB Simulation.

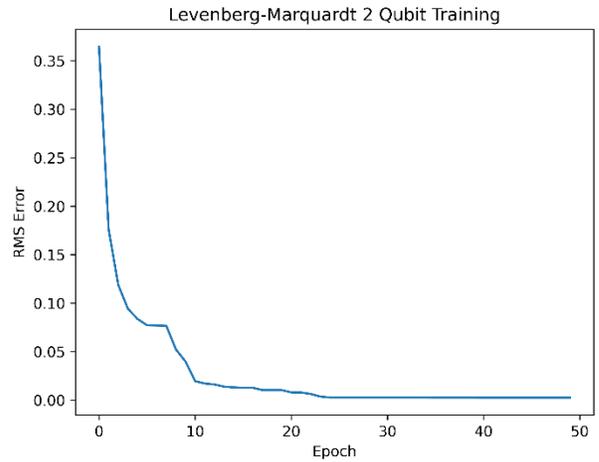

Fig. 10. RMS error vs. epoch for Levenberg-Marquardt 2-qubit training in Qiskit.

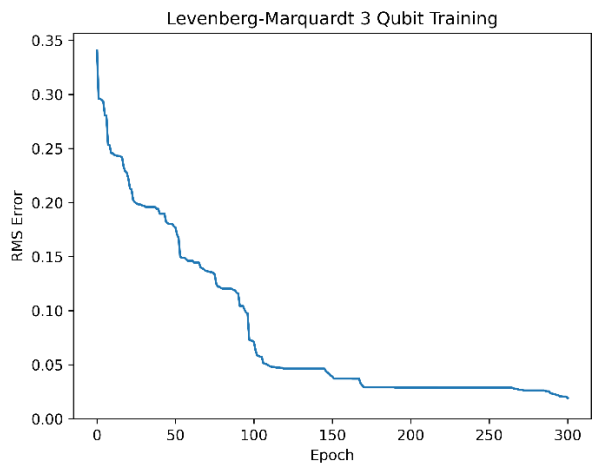

Fig. 11 RMS error vs. epoch for Levenberg-Marquardt 3-qubit training in Qiskit.

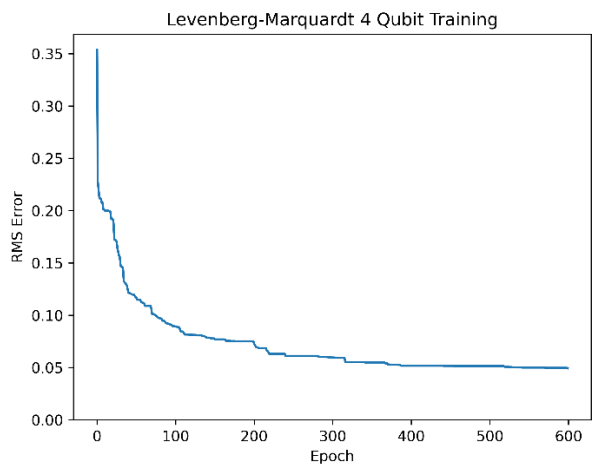

Fig. 12 RMS error vs. epoch for Levenberg-Marquardt 4-qubit training in Qiskit.

Acknowledgment

We all thank the entire research group for helpful discussions: Nam Nguyen, Saideep Nannapaneni, William Ingle, Henry Elliott, Ricardo Rodriguez, and Sima Borujeni.

VI. References

[1] E. C. Behrman, L. R. Nash, J. E. Steck, V. G. Chandrashekar and S. R. Skinner, "Simulations of quantum neural networks," *Information Sciences,* vol. 128, pp. 257-269, 2000.

[2] E. C. Behrman, J. E. Steck, P. Kumar and K. A. Walsh, "Quantum algorithm design using dynamic learning," *Quantum Information & Computation,* vol. 8, pp. 12-29, 2008.

[3] E. C. Behrman and J. E. Steck, "Multiqubit entanglement of a general input state," *Quantum Information & Computation,* vol. 13, pp. 36-53, 2013.

[4] N. H. Nguyen, E. C. Behrman and J. E. Steck, "Quantum learning with noise and decoherence: a robust quantum neural network," *Quantum Machine Intelligence,* vol. 2, pp. 1-15, 2020.

[5] N. H. Nguyen, E. C. Behrman, M. A. Moustafa and J. E. Steck, "Benchmarking neural networks for quantum computations," *IEEE Transactions of Neural Networks and Learning Systems,* vol. 31, pp. 2522-2531, 2020.

[6] M. Caro, H.-Y. Huang, M. Cerezo, K. Sharma, A. Sornborger, L. Cincio and P. Coles, "Generalization in quantum machine learning from few data," *Nature Communications,* pp. https://doi.org/10.1038/s41467-022-32550-3, 2022.

[7] L. Gurvitz, "Classical deterministic complexity of Edmonds problem and quantum entanglement," in *Proceedings of the 35th Annual ACM Symposium on Theory of Computing*, 2003.

[8] J. Preskill, "Quantum entanglement and quasntum computing," in *Proceedings of the 25th Solvey Conference on Physics*, 2013.

[9] P. J. Werbos, "Neurocontrol and supervised learning: An overview and evaluation," in *Handbook of Intelligent Control*, 1992.

[10] C. Goncalves, "Quantum neural machine learning: Backpropagation and dynamics," *NeuroQuantology,* vol. 15, pp. 22-41, 2017.

[11] G. Verdun, J. Pye and M. Broughton, "A universal training algorithm for quantum deep learning," 2018. [Online]. Available: arXiv:1806.09729.

[12] H. Abraham and et al., "Qiskit: An open source framework for quantum computing," 2019.

[13] N. L. Thompson, J. E. Steck and E. C. Behrman, "A non-algorithmic approach to "programming" quantum computers via machine learning," in *IEEE International Conference on Quantum Computing and Engineering*, 2020.

[14] R. P. Feynman, "An operator calculus having applications in quantum electrodynamics," *Physical Review,* vol. 84, pp. 108-128, 1951.

[15] J. J. Sakurai, Modern Quantum Mechanics, 2017.

[16] N. L. Thompson, N. H. Nguyen, E. C. Behrman and J. E. Steck, "Experimental pairwise entanglement estimation for an N-qubit system: A machine learning approach for programming quantum hardware," *Quantum Information Processing,* vol. 19, pp. 1-18, 2020.

[17] E. C. Behrman, N. H. Nguyen, J. E. Steck and M. McCann, "Quantum neural computation of entanglement is robust to noise and decoherence," in *Quantum Inspired Computational Intelligence*, Boston, Morgan-Kauffmann, 2017, pp. 3-32.

[18] K. Levenberg, "A method for the solution of certain non-linear problems in least squares," *Quarterly of Applied Mathematics*, vol. 2, pp. 164-168, 1944.

[19] D. W. Marquardt, "An algorithm for least-squares estimation of nonlinear parameters," *Journal of the Society for Industrial and Applied Mathematics,* vol. 11, pp. 431-441, 1963.

[20] M. K. Transtrum and J. P. Sethna, "Improvements to the Levenberg Marquardt algorithm for nonlinear least-squares minimization," 2012. [Online]. Available: arXiv:1201.5885.

[21] G. E. Crooks, "Gradients of parametrized quantum gates using the parameter shift rule and gate decomposition," 2019. [Online]. Available: arXiv:1905.1311.

[22] M. K. Transtrum, B. B. Machta and J. P. Sethna, "Geometry of nonlinear least squares with applications to sloppy models and optimization," *Physical Review E,* vol. 83, p. 036701, 2011.

[23] J. R. McClean, S. Boixo, V. N. Smelyanskiy, R. Babbush and H. Neven, "Barren plateaus in quantum neural network training landscapes," *Nature Communications,* vol. 9, p. 4812, 2018.

[24] John Towns, Timothy Cockerill, Maytal Dahan, Ian Foster, Kelly Gaither, Andrew Grimshaw, Victor Hazlewood, Scott Lathrop, Dave Lifka, Gregory D. Peterson, Ralph Roskies, J. Ray Scott, Nancy Wilkins-Diehr, ""XSEDE: Accelerating Scientific Discovery",," *Computing in Science & Engineering,* vol. 16, no. doi:10.1109/MCSE.2014.80, pp. 62-74, 2014.

[25] P. Shor, "Algorithms for quantum computation: discrete logarithms and factoring," in *Proceedings of the 35th Annual Symposium on Foundations of Computer Science*, 1994.

[26] L. Grover, "A fast quantum mechanical algorithm for database search," in *Proceedings of the 28th Annual ACM Symposium on Theory of Computing*, 1996.

[27] A. M. Childs, R. Cleve, E. Deotto, E. Farhi, S. Gutman and D. A. Spielman, "Exponential algorithmic speedup by a quantum walk," in *Proceedings of the 35th Annual ACM Symposium on Theory of Computing*, 2003.

[28] S. Bravyi, D. Gosset and R. Konig, "Quantum advantage with shallow circuits," *Science,* vol. 362, pp. 308-311, 2018.

[29] T. F. Ronnow, Z. Wang, J. Job, S. Boixo, S. V. Isakov, D. Wecker, J. M. Martinis, D. A. Lidar and M. Troyer, "Defining and detecting quantum speedup," *Science,* vol. 345, pp. 420-424, 2014.